\documentstyle[12pt,amssymb,amsfonts,amsbsy]{article}
\begin{document}
\author{Ilja Schmelzer\thanks
       {WIAS Berlin}}

 \title{A Metric Theory of Gravity with Condensed Matter Interpretation}

\maketitle

\begin{abstract}
\sloppypar We consider a classical condensed matter theory in a
Newtonian framework where conservation laws

\begin{eqnarray*}
\partial_t \rho + \partial_i (\rho v^i) &= &0 \\
\partial_t (\rho v^j) + \partial_i(\rho v^i v^j + p^{ij}) &= &0
\end{eqnarray*}

are related with the Lagrange formalism in a natural way.  For an
``effective Lorentz metric'' $g_{\mu\nu}$ it is equivalent to a metric
theory of gravity close to general relativity with Lagrangian

\[
L = L_{GR}
  - (8\pi G)^{-1}(\Upsilon g^{00}-\Xi (g^{11}+g^{22}+g^{33}))\sqrt{-g}
\]

We consider the differences between this theory and general relativity
(no nontrivial topologies, stable frozen stars instead of black holes,
big bounce instead of big bang singularity, a dark matter term),
quantum gravity, and the connection with realism and Bohmian mechanics.

\end{abstract}

\section{Introduction}

General relativity is a very beautiful and successful theory of
gravity.  Nonetheless, the consideration of alternative theories of
gravity remains to be legitimate part of science.  Even if the purpose
is only ``to play devil's advocate'', as in the case of some
interesting theories of gravity (Lightman and Lee \cite{Lightman}, Ni
\cite{Ni}) or to answer Rosen's \cite{Rosen} question ``whether one
can set up a theory of gravitation which will give agreement with
observation without permitting black holes''.

The current research was motivated by the conceptual conflict between
general relativity and quantum theory.  The basic idea was that the
Einstein equations may appear in a completely different metaphysical
framework, which is better compatible with quantum principles than
relativistic spacetime. This idea is in itself in full agreement with
``the present educated view on the standard model, and of general
relativity, ... that these are leading terms in effective field
theories'' \cite{Weinberg}.  The simplest choice would be a classical
Newtonian framework with absolute time.  This is a known way to solve
the ``problem of time'' in quantum gravity, usually rejected for
metaphysical reasons: ``... in quantum gravity, one response to the
problem of time is to `blame' it on general relativity's allowing
arbitrary foliations of spacetime; and then to postulate a preferred
frame of spacetime with respect to which quantum theory should be
written.  Most general relativists feel this response is too radical
to countenance: they regard foliation-independence as an undeniable
insight of relativity.''  \cite{Butterfield}.

Nonetheless, following this ``too radical to countenance'' way, we
have found a surprisingly simple and beautiful scheme which allows to
{\em derive} a variant of the Einstein equations based not only on the
classical Newtonian framework, but also on classical condensed matter
theory -- in other words, an ideal realization of the last century
``ether'' concept.  In this derivation, we do not need any conspiracy
to explain the Einstein equivalence principle.  All we need are
classical conservation laws and their connection with the Lagrange
formalism.  The point is the combination of the symmetry of the
Lagrange formalism (self-adjoint equations) with the special character
of the conservation laws (their relation with the preferred
coordinates).

The mere existence of a viable theory of gravity with preferred frame
is of great importance for other foundational problems.  A preferred
frame is, for example, required for compatibility of the EPR criterion
of reality \cite{EPR} with the violation of Bell's inequality
\cite{Aspect}.  Bell himself concludes \cite{Bell1}: ``the cheapest
resolution is something like going back to relativity as it was before
Einstein, when people like Lorentz and Poincare thought that there was
an aether --- a preferred frame of reference --- but that our
measuring instruments were distorted by motion in such a way that we
could no detect motion through the aether.''  The theory presented
here is strong support for this ``cheapest resolution''. A closely
related question is the extension of Bohmian mechanics \cite{Bohm}
into the domain of relativistic gravity which requires a preferred
frame too.

The resulting theory differs from general relativity in an interesting
way.  It contains additional terms which depend on the preferred
frame.  These additional terms allow the definition of local energy
and momentum densities of the gravitational field.  But they don't
violate the Einstein equivalence principle -- the theory remains to be
a metric theory of gravity.  They influence only the gravitational
field itself, similar to dark matter.

The close analogy between condensed matter theory and gravity is
well-known.  It has been recognized that ``effective gravity, as a
low-frequency phenomenon, arises in many condensed matter systems''
\cite{Volovik}.  This has been used to study Hawking radiation and the
Unruh effect \cite{Visser} \cite{Unruh} \cite{Jacobson} \cite{Volovik}
and vacuum energy \cite{Volovik} for condensed matter examples.
Wilczek \cite{Wilczek} mentions the general exchange of ideas with
high energy physics, which ``includes global and local spontaneous
symmetry breaking, the renormalization group, effective field theory,
solitons, instantons, and fractional charge and statistics''.  This
analogy has suggested the idea of a ``Planck ether''
\cite{Jegerlehner}.  Our theory fits very well into this general
context, and suggests interesting modifications: the critical length
should not be Planck length.

\section{The Theory}

Our theory describes a classical medium in a Newtonian framework --
Euclidean space and absolute time.  The medium is described by steps
of freedom typical for condensed matter theory.  The gravitational
field is defined by a positive density $\rho$, a velocity $v^i$, and a
negative-definite symmetrical tensor field $p^{ij}$ which we name
``pressure''.  The effective metric $g_{\mu\nu}$ is defined
algebraically by

\begin{eqnarray*}\label{gdef}
 \hat{g}^{00} = g^{00} \sqrt{-g} &=  &\rho \\
 \hat{g}^{i0} = g^{i0} \sqrt{-g} &=  &\rho v^i \\
 \hat{g}^{ij} = g^{ij} \sqrt{-g} &=  &\rho v^i v^j + p^{ij}
\end{eqnarray*}

This decomposition of $g^{\mu\nu}$ into $\rho$, $v^i$ and $p^{ij}$ is
a variant of the ADM decomposition.  The signature of $g^{\mu\nu}$
follows from $\rho>0$ and negative definiteness of $p^{ij}$.

The theory does not specify all properties of the medium, but only a
few general properties -- the conservation laws and their relation to
the Lagrange formalism.  The ``material properties'' of the medium,
denoted by $\varphi^m$, remain unspecified.  They become the matter
fields.  The complete specification of the medium -- which includes
the material laws of the medium -- gives the theory of everything.
The few general properties fixed here define a theory of gravity
similar to GR.  While it leaves the matter steps of freedom and the
matter Lagrangian unspecified, it derives the Einstein equivalence
principle.

For the derivation of the Lagrange formalism we prefer a formalism
where the non-covariant terms are disguised as covariant, with the
preferred coordinates considered formally as scalar fields $X^\mu(x)$.
\footnote{It is well-known that every physical theory may be described
in a covariant way.  But usually this is done in another way (for
example, by Fock \cite{Fock} for SR): a flat background metric
$\gamma_{\mu\nu}$ is described by vanishing curvature
$R^\mu_{\nu\kappa\lambda}=0$.}  It is easy to transform a
non-covariant Lagrangian $L=L(T^{\ldots}_{\ldots},\partial_\mu
T^{\ldots}_{\ldots})$ into a (formally) covariant form
$L=L(T^{\ldots}_{\ldots},\partial_{\mu}T^{\ldots}_{\ldots},X^\mu_{,\nu})$.
For example, the non-covariant component $a^0$ of a vector $a^\mu$
will be written as $a^\nu X^0_{,\nu}$.  Here $X^0 = T$ is no longer a
spatial index, but enumerates one of the four ``scalar fields''
$T,X,Y,Z$.

This formalism is interesting in itself.  Especially the question how
to distinguish ``truly covariant'' theories from theories made
covariant using this formalism is a very interesting one.  The most
interesting point of the formalism is the relation between
conservation laws and the preferred coordinates.  The conservation
laws may be defined as the Euler-Lagrange equations for the preferred
coordinates.  The related energy-momentum tensor

\[ T^\nu_\mu = - {\partial L\over\partial X^\mu_{,\nu}} \]

is not the same as in Noether's theorem, but is equivalent: if the
Lagrangian does not depend on the $X^\mu$ them-self, we obtain
immediately conservation laws in the form

\[ \partial_\nu T^\nu_\mu = 0 \]

Now, the main postulate of the theory is that these conservation laws
are identified with the classical conservation laws we know from
condensed matter theory.  First, the Euler-Lagrange equation for the
preferred time T we identify with the classical continuity equation
for the medium:

\begin{equation} \label{continuity}
 \partial_t \rho + \partial_i (\rho v^i) = 0
\end{equation}

The equations for the preferred spatial coordinates $X^i$ we identify
with the Euler equation:

\begin{equation} \label{momentum}
 \partial_t (\rho v^j) + \partial_i(\rho v^i v^j+p^{ij})=0
\end{equation}

Note that the Euler equation contains an important physical
assumption: there is no momentum exchange with other materials,
because there are no other materials.  We have only one, universal,
medium.  All usual ``matter fields'' $\varphi^m$ are material
properties of this universal medium.

The four conservation laws transform into the harmonic condition for
the metric $g_{\mu\nu}$.  Thus, they really look like equations for
the preferred coordinates:

\[ \Box X^\nu = \partial_\mu (g^{\mu\nu}\sqrt{-g}) = 0\]

Therefore, the main postulate transforms into the following relation
between the the Euler-Lagrange equations for $S = \int L$ and the
preferred coordinates $X^\mu$:

\[{\delta S\over\delta X^\mu}\equiv-(4\pi G)^{-1}\gamma_{\mu\nu}\Box X^\nu\]

We have introduced here a constant diagonal matrix $\gamma_{\mu\nu}$
and a common factor $-(4\pi G)^{-1}$ to obtain appropriate units
below.  Euclidean symmetry gives
$\gamma_{11}=\gamma_{22}=\gamma_{33}$.  Thus, we have two coefficients
$\gamma_{00}=\Upsilon,\gamma_{ii}=-\Xi$.  Now, we can derive the
general form of the Lagrangian.  First, we have the particular
Lagrangian

\[L_{0} = -(8\pi G)^{-1}\gamma_{\mu\nu}X^\mu_{,\alpha}X^\nu_{,\beta}
      	g^{\alpha\beta}\sqrt{-g}\]

which fulfils this property.  For the difference $L-L_{0}$ we obtain

\[ {\delta \int(L-L_{0})\over\delta X^\mu} \equiv 0 \]

Thus, the remaining part is not only covariant in the weak, formal
sense enforced by our decision to handle the preferred coordinates as
fields.  It does not depend on the preferred coordinates $X^\mu$.  But
this is the original ``strong'' covariance, the classical requirement
for the Lagrangian of general relativity.  Thus, we can identify the
difference $L-L_{0}$ with the most general classical Lagrangian of
general relativity.  In the preferred coordinates we obtain

\[L = -(8\pi G)^{-1}\gamma_{\mu\nu}g^{\mu\nu}\sqrt{-g}
    + L_{GR}(g_{\mu\nu}) + L_{matter}(g_{\mu\nu},\varphi^m).
\]

Note that this Lagrangian fulfils the Einstein equivalence principle
in its full beauty.  That means, we have derived this principle
starting with few general assumptions about a medium in a classical
Newtonian framework.

This derivation of exact relativistic symmetry in the context of a
classical condensed matter theory is the main result of this paper.
To improve our understanding, let's consider how this has happened,
and what has been really used to derive the EEP.  The derivation is
extremely simple, but given in an unusual formalism.

But how relativistic symmetry appears may be explained without
reference to this formalism.  There are three principles involved:
first, the inherent symmetry of the Lagrange formalism -- the
equations should be self-adjoint, or ``action equals reaction''.
Next, there is the relation between preferred coordinates and
conservation laws in the Lagrange formalism well-known from Noether's
theorem.  And, last not least, we have the independence of the
conservation laws from the material properties of the medium enforced
by our choice of variables.  As a consequence of these principles, the
Euler-Lagrange equations for the material properties do not depend on
the preferred coordinates.  But this is already the EEP:

\[ {\delta\over\delta X^\mu}{\delta S \over\delta \phi^m}
= {\delta \over\delta \phi^m}{\delta S \over\delta X^\mu}
= {\delta \over\delta \phi^m} \mbox{[cons. laws]} = 0 \]

The following heuristic, informal picture may be useful for the
understanding: The medium is universal.  All usual matter fields
(gauge fields, fermions) are material properties of this medium,
something like defect densities in a crystal.  In this picture human
beings consist of crystal defects and interact only with other crystal
defects.  It seems quite obvious that such beings have only restricted
observational possibilities.  This restriction of observational
possibilities leads to relativistic symmetry -- we cannot distinguish
by observation states which are really different.  Thus, there is
nothing strange in the appearance of relativistic symmetry in usual
condensed matter.  It is a natural consequence of the special nature
of matter fields in this theory -- they are all material properties of
a single universal medium.

\subsection{Equations and Energy-Momentum Tensor}

After the derivation of the theory, the ``covariant formalism'' has
done its job, and we can return to a form more appropriate for the
comparison with other theories of gravity.  We obtain the following
equations:

\[G^\mu_\nu  = 8\pi G (T_m)^\mu_\nu
   + (\Lambda +\gamma_{\kappa\lambda}g^{\kappa\lambda}) \delta^\mu_\nu
   - 2g^{\mu\kappa}\gamma_{\kappa\nu}.\]

The harmonic condition

\[ \partial_\mu  (g^{\mu\kappa}\sqrt{-g}) = 0 \]

is a consequence of these equations and one form of energy-momentum
conservation in the theory.  Remarkably, there is also another form --
the basic equation may be simply considered as a decomposition of the
full energy-momentum tensor $g^{\mu\kappa}\sqrt{-g}$ into a part which
depends on matter fields and a part which depends on the gravitational
field:

\[(T_g)^\mu_\nu = (8\pi G)^{-1}\left(\delta^\mu_\nu(\Lambda
              	+ \gamma_{\kappa\lambda}g^{\kappa\lambda})
	 	- G^\mu_\nu\right)\sqrt{-g}\]

Thus, instead of no local conservation law in GR we obtain even two
equivalent forms of local conservation laws.  The first is equivalent
to classical conservation laws from condensed matter theory and, in
our variables, to the harmonic condition.  The other is equivalent to
the conservation law in Noether's theorem, and splits into a part
which depends on the ``matter fields'' $\varphi^m$ and a purely
gravitational part.

\section{Predictions}

Using small enough values $\Xi, \Upsilon\to 0$ leads to the classical
Einstein equations.  Therefore it is not problematic to fit
observation.  It is much more problematic to find a way to distinguish
our theory from GR by observation.

\subsection{A dark matter candidate}

Let's consider the influence of the new terms on the expansion of the
universe.  In our theory a homogeneous universe should be flat.
Solutions with non-zero curvature may be solutions of our theory too,
but they cannot be homogeneous.  The the usual ansatz $ds^2 = d\tau^2
- a^2(\tau)(dx^2+dy^2+dz^2)$ gives

\begin{eqnarray*}
3(\dot{a}/a)^2  &=&
   	- \Upsilon/a^6 + 3 \Xi/a^2 + \Lambda + \varepsilon\\
2(\ddot{a}/a) + (\dot{a}/a)^2 &=&
   	+ \Upsilon/a^6 +   \Xi/a^2 + \Lambda - p
\end{eqnarray*}

We see that $\Xi$ influences the expansion of the universe similar to
homogeneous (hot) dark matter with $p=-{1\over 3}\varepsilon$.

\subsection{Big bounce instead of big bang singularity}

$\Upsilon$ becomes important only in the very early universe.  But for
$\Upsilon>0$, we obtain a qualitatively different picture. We obtain a
lower bound $a_0$ for $a(\tau)$ defined by

\[ \Upsilon/a_0^6 = 3 \Xi/a_0^2 + \Lambda + \varepsilon \]

The solution becomes symmetrical in time, with a big crash followed by
a big bang.  For example, if $\varepsilon = \Xi =0, \Upsilon>0,
\Lambda>0$ we have the solution

\[ a(\tau) = a_0 \cosh^{1/3}(\sqrt{3\Lambda} \tau)  \]

In time-symmetrical solutions of this type the horizon is, if not
infinite, at least big enough to solve the cosmological horizon
problem (cf. \cite{Primack}) without inflation.

\subsection{Frozen stars instead of black holes}

The choice $\Upsilon>0$ influences also another physically interesting
solution -- the gravitational collapse.  There are stable ``frozen
star'' solutions with radius slightly greater than their Schwarzschild
radius.  The collapse does not lead to horizon formation, but to a
bounce from the Schwarzschild radius.  Let's consider an example.  The
general stable spherically symmetric harmonic metric depends on one
step of freedom m(r) and has the form

\[ ds^2 = (1-{m \over r}{\partial m\over\partial r})
      	     	({r-m\over r+m}dt^2-{r+m\over r-m}dr^2)
       	- (r+m)^2 d\Omega^2 \]

Let's consider the ansatz $m(r)=(1-\Delta)r$. We obtain

\begin{eqnarray*}
ds^2 &=& \Delta^2dt^2 - (2-\Delta)^2(dr^2+r^2d\Omega^2) \\
0    &=& -\Upsilon \Delta^{-2} +3\Xi(2-\Delta)^{-2}+\Lambda+\varepsilon\\
0    &=& +\Upsilon \Delta^{-2} + \Xi(2-\Delta)^{-2}+\Lambda-p
\end{eqnarray*}

Now, for very small $\Delta$ even a very small $\Upsilon$ becomes
important, and we obtain a non-trivial stable solution for $p =
\varepsilon = \Upsilon g^{00}$.  Thus, the surface remains visible,
with time dilation $\sqrt{\varepsilon/\Upsilon}\sim M^{-1}$.

\section{Relativity Principle, Realism and Bohmian mechanics}

As we have shown, we have relativistic symmetry for all observable
effects (relativity principle for observables).  On the other hand,
reality itself does not have this relativistic symmetry (no relativity
principle for reality).

But this distinction is meaningful only if we have a realistic theory.
Here we define realism in the following classical way: Assume we have
an experiment described by observables $X$ with the observable
probability distribution $\rho_X(X,x)dX$, which depends on a set of
control parameters $x$ (the decisions of experimenters).  In a
realistic theory this is described by some reality $\lambda\in\Lambda$
with observer-independent probability distribution
$\rho_\lambda(\lambda)d\lambda$ which explains the observations X
with a function $X(x,\lambda)$ so that for a test function f we have:

\[ \int f(X)\rho_X(X,x)dX=\int f(X(x,\lambda))\rho(\lambda) d\lambda \]

This definition is appropriate to define causality in a natural way:
The decision of the experimenter x has a causal influence on X if this
function $X(x,\lambda)$  depends on x.

These definitions are sufficient to proof Bell's inequality following
\cite{Bell}.  Therefore, the violation of Bell's inequality
\cite{Aspect} defines a contradiction between realism, causality and
the relativity principle for reality.  If we want to preserve the full
relativity principle, we have to reject realism or causality, and the
usual decision is the rejection of realism.  But in our theory the
relativity principle is automatically restricted to observable
effects.  Therefore, no contradiction appears.  We cannot prove Bell's
inequality for space-like separated events and, therefore, have no
conflict with Aspect's experiment.  This solution of the puzzle has
been preferred by Bell \cite{Bell1}: ``the cheapest resolution is
something like going back to relativity as it was before Einstein,
when people like Lorentz and Poincare thought that there was an aether
--- a preferred frame of reference --- but that our measuring
instruments were distorted by motion in such a way that we could not
detect motion through the aether. Now, in that way you can imagine
that there is a preferred frame of reference, and in this preferred
frame of reference things go faster than light.''

In this context, Bohmian mechanics (BM) \cite{Bohm} is very important.
BM is a realistic, even deterministic, theory which makes in a
``quantum equilibrium'' the same predictions as quantum theory.  In
the relativistic context it requires a preferred frame.  This is
usually considered as a decisive argument against BM.  But in the
context of our theory this argument no longer holds: our theory shows
the way how to extend BM into the domain of relativistic gravity.

Let's add a few additional arguments in favour of realism: In our
definition, realism is not related with a particular spacetime theory,
it does not even depend on the notion of spacetime.  This makes
realism more fundamental than spacetime theory, and, moreover, more
fundamental than a particular spacetime theory like relativity.  In
the case of conflict, the natural decision is to reject the less
fundamental theory.  In our case, it is the relativity principle for
reality, and not realism.  Moreover, there is no independent evidence
against realism -- this is proven by the existence of BM for all
quantum effects.  Instead, the problems related with quantum theory,
especially the problem of time \cite{Isham}, may be interpreted as
independent evidence against relativity.  And, last not least, we
loose essentially nothing if we reduce the relativity principle to
observable effects.  If we reject realism, the relativity principle
essentially reduces to observable effects too.

As we see, the relation between our theory, realism, and BM is mutual
support.  On one hand, compatibility with realism and BM are strong
arguments in favour of our theory.  On the other hand, our theory
weakens the most serious arguments against realism and BM --
incompatibility with relativistic principles.

\section{Quantization}

Most workers would agree that ``at the root of most of the conceptual
problems of quantum gravity'' is the idea that ``a theory of quantum
gravity must have something to say about the quantum nature of space
and time'' \cite{Butterfield}.  These problems obviously disappear for
a theory of gravity with fixed Newtonian background.  Especially this
holds for the ``problem of time'': ``... in quantum gravity, one
response to the problem of time is to `blame' it on general
relativity's allowing arbitrary foliations of spacetime; and then to
postulate a preferred frame of spacetime with respect to which quantum
theory should be written.'' \cite{Butterfield}.  In this way, in our
theory the problem of time simply disappears.  Now, ``most general
relativists feel this response is too radical to countenance: they
regard foliation-independence as an undeniable insight of
relativity.''  \cite{Butterfield}.  That means, this rejection is
based on metaphysical preference for the GR spacetime concept only.

It should be noted that together with black holes another quantization
problem disappears -- the information loss problem.  The frozen stars
are stable and do not evaporate.  Most problems related with energy
and momentum conservation disappear too -- the Hamiltonian is no
longer a constraint.

To solve the ultraviolet problems, we have to make an additional
assumption, but a very natural one for a condensed matter theory: an
``atomic hypothesis''.  After this, the theory is in ideal agreement
with ``the present educated view on the standard model, and of general
relativity, ... that these are leading terms in effective field
theories'' \cite{Weinberg} -- an idea introduced by Sakharov
\cite{Sakharov}.  An interpretation of $\rho$ as the number of
``atoms'' per volume leads to an interesting prediction for the
cutoff:

\[ \rho(x) V_{cutoff} = 1. \]

It is non-covariant.  For a homogeneous ``expanding'' universe, it
seems to expand together with the universe.  Thus, the cutoff differs
in a principal way from the usual expectation that the cutoff is the
Planck length $a_P\approx 10^{-33}cm$ (cf. \cite{Jegerlehner},
\cite{Volovik}).

\section{Comparison with other theories of gravity}

Because of the simplicity of the additional terms it is no wonder that
they have been already considered.  Two other theories have a similar
Lagrangian for appropriate signs of the cosmological constants: the
``relativistic theory of gravity'' proposed by Logunov et
al. \cite{Logunov} and classical GR with some additional scalar ``dark
matter'' fields.  Nonetheless, equations are not all.  There are other
physical important things which makes the theories different as
physical theories, like global restrictions, boundary conditions,
causality restrictions, quantization concepts which are closely
related with the underlying ``metaphysical'' assumptions.

\subsection{Comparison with RTG}

The ``relativistic theory of gravity'' (RTG) proposed by Logunov et
al. \cite{Logunov} has Minkowski background metric $\gamma_{\mu\nu}$.
The Lagrangian of RTG is

\[L = L_{Rosen} + L_{matter}(g_{\mu\nu},\psi^m)
    - {m_g^2}({1\over 2}\gamma_{\mu\nu}g^{\mu\nu}\sqrt{-g}
            - \sqrt{-g} - \sqrt{-\gamma})
\]

which de facto coincides with our theory for $\Lambda = - {m_g^2} <
0$, $\Xi = - \gamma^{11}{m_g^2} > 0$, $\Upsilon = \gamma^{00}{m_g^2} > 0$.

The metaphysical concept of RTG is completely different.  It is a
special-relativistic theory, therefore incompatible with classical
realism and Bohmian mechanics because of the violation of Bell's
inequality.  Another difference is the causality condition: In RTG,
only solutions where the light cone of $g_{ij}$ is inside the light
cone of $\gamma_{ij}$ are allowed.  A comparable but weaker condition
exists in our theory too: $T(x)$ should be a time-like function, or,
$\rho(X,T)>0$.

The metaphysical differences become physical if we consider
quantization.  Indeed, RTG suggests quantization following standard
QFT schemes.  Instead, our theory suggests to quantize an atomic
model.  These ways are conceptually incompatible.  Indeed, the
prediction for the cutoff length $l_{cutoff}$, which is based on the
interpretation of $g^{00}\sqrt{-g}$ as the atomic density, is not
Lorentz-covariant and therefore incompatible with RTG.

\subsection{Comparison with GR plus scalar fields}

In the formalism where the preferred coordinates are handled as scalar
fields, the Lagrangian of our theory looks equivalent to GR with some
dark matter -- four scalar fields $X^{\mu}$.  Similar ``clock fields''
in GR have been considered by Kuchar \cite{Kuchar}.  Usual energy
conditions require $\Xi>0, \Upsilon<0$.  To obtain the most
interesting effects (no black holes, no big bang singularity) we have
to choose $\Upsilon>0$, which violates the usual GR energy conditions.

Nonetheless, even if the Lagrangian seems to be the same, the theories
are completely different as physical theories.  In a typical solution
of GR with scalar fields the fields $X^\mu(x)$ cannot be used as
global coordinates.  Especially this holds for all solutions with
non-trivial topology.  Even if they may be used as global coordinates,
the field $T(x)$ may be not time-like.  All these solutions are
forbidden in our theory. \footnote{That means, their observation
falsifies our theory, but does not falsify GR with scalar fields.
According to Popper's criterion of empirical content this means higher
empirical content for our theory.}

But the remaining solutions -- that means, solutions of our theory
which may be interpreted as solutions of GR with scalar fields too --
are very unnatural from point of this theory.  They have very strange
boundary values -- the fields $X^\mu(x)$ are unbounded.  Thus, if we
consider boundary conditions of type $|X^\mu(x)|<C$ as part of GR with
scalar fields, then we have no common solutions for above theories.

As we see, to handle preferred coordinates like scalar fields is
justified only in a very restricted domain.  It does not make our
condensed matter theory on a preferred background equivalent to a
general-relativistic theory, even if the Lagrangian has the same form.
Preferred coordinates and scalar fields remain to be very different
things.

Of course, above theories differ also in their metaphysical principles
and their quantization concepts.  Especially it is incompatible with
classical realism and Bohmian mechanics.  Similarly, we have a
completely different quantization approach, the cutoff length
$l_{cutoff}$ is not Lorentz-covariant and therefore incompatible with
GR metaphysics.  In our theory the $X^\mu$ are not fields, but fixed
background coordinates and therefore should not be quantized, while
the ``fields'' $X^\mu(x)$ in GR with scalar fields should be
quantized.

\section{Conclusions}

We have started with postulates for a medium in a classical Newtonian
world: classical conservation laws and their connection with the
Lagrange formalism.  We have obtained a viable theory of gravity
which, in a certain limit $\Xi,\Upsilon\to 0$, leads to the classical
Einstein equations.  We have derived the Einstein equivalence
principle from these first principles.

The resulting theory is compatible with classical realism and Bohmian
mechanics.  This is not only an argument in favour of this theory, but
removes serious arguments against realism and Bohmian mechanics,
making them viable in the domain of relativistic gravity.

The theory has a lot of other interesting advantages in comparison
with general relativity: well-defined local energy and momentum
densities, a classical Hamilton formalism, no black hole and big bang
singularities, no cosmological horizon problem, a natural dark matter
candidate, no problem of time in quantum gravity, no information loss
problem, no problems with non-trivial topologies, a natural ``atomic
ether'' quantization concept compatible with modern effective field
theory.

Are there serious disadvantages?  The relativity principle is
restricted to observable effects, but the rejection of realism related
with relativistic quantum theory has a similar effect in relativity.
SF authors probably don't like that non-trivial topologies and causal
loops are forbidden.  Nonetheless, it is an advantage according to
Popper's criterion of empirical content.  Is our theory less beautiful
than GR?  That's, of course, a matter of taste.  But many beautiful
aspects of GR appear in our theory too, and some very beautiful
concepts often used but of no fundamental importance in GR (ADM
decomposition, harmonic gauge) play a fundamental role in our theory.
The situation with conservation laws is certainly more beautiful in
our theory.

But, even if you nonetheless decide to prefer relativistic theory --
the mere existence of a theory which, based on first principles,
predicts a variant of the Einstein equations is an interesting fact.
As a consequence, the strong empirical evidence in favour of the
Einstein equations them-self (Solar system observations, binary
pulsars) are no longer support for general-relativistic spacetime
concepts.  The choice between relativistic spacetime and classical
ether should be justified in a different way.

\end{document}